\documentclass[epj]{svjour}
\usepackage{graphicx}
\begin{document}
\title{In-medium effects on electromagnetic probes}
\author{Charles Gale
%\thanks{\emph{Present address:} Service de Physique Th\'eorique}%
}                     % Do not remove
\institute{McGill University, Department of Physics, 3600 University Street, 
Montreal QC, Canada H3A 2T8}

\date{Received: date / Revised version: date}
\abstract{
We discuss some of the aspects of the physics of relativistic nuclear 
collisions, 
in particular those having to do with the observation of electromagnetic
radiation. We concentrate on what such measurements tell us about the local, 
in-medium properties of the environment from which they emerge. The
contribution from different sources are considered: that from the partonic
sector of QCD, and that from the confined hadronic phase. Specifically, we
discuss the observation of real photons and of lepton pairs at the SPS and at 
RHIC,
and make predictions for the LHC. The role of jets is discussed.
\PACS{
      {25.75.-q}{Relativistic heavy-ion collisions}   \and
      {12.38.Mh}{Quark-gluon plasma}  
     } % end of PACS codes
} %end of abstract
\maketitle
\section{Introduction}
\label{intro}

Electromagnetic radiation defines a privileged class of observables in the
study of relativistic nuclear collisions. As real and virtual photons are
only weakly coupled (in a parametrical sense) to the strongly interacting 
medium they are excellent probes of the local conditions at the time of their
emission, because of the absence of final-state interaction. Of course, the 
physical interpretation of the information carried by such measurements also 
requires a knowledge of the space-time evolution of the emitting medium. With
those aspects in mind, we first recall the results of low-mass dilepton
measurements at the SPS. We reiterate that those results are consistent with an
interpretation in terms of vector spectral densities that are different from
what they are in vacuum. We then show that those same spectral densities can
be used to theoretically interpret the real photon spectrum, also measured at
the SPS. Together with the intermediate invariant mass regime, we conclude
that the case for the observation of a new phase of QCD through
electromagnetic measurements at SPS energies, even though suggestive, 
can't be made convincingly. At still higher energies, photon
production and jet quenching are considered consistently through jet-plasma
interactions. We point out that this new source has even been seen at RHIC.

\section{Low-mass lepton pairs}

\begin{figure}[!th]
\begin{center}
\resizebox{0.444\textwidth}{!}{%
  \includegraphics[angle=0]{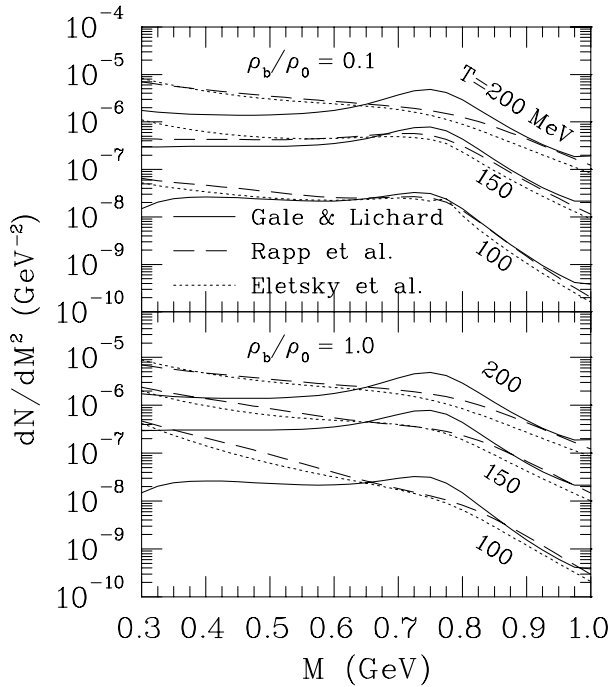}
}
\end{center}
%\vspace{5cm}       % Give the correct figure height in cm
\caption{A comparison \cite{pasi} of the vector spectral densities, obtained in the
effective Lagrangian approach \cite{RaWa,raga}, and through a direct
experimental fit of the scattering amplitudes \cite{ebek}. Also shown as a
baseline is the result consisting of an incoherent sum of baryon-free channels
\cite{gali}.}
\label{rate_comp}       % Give a unique label
\end{figure}

At SPS energies, the measurement of low-mass lepton pairs has first 
been made by the
Helios/3 \cite{helios3}, and then by the CERES \cite{ceres} 
collaborations. Those latter data represent the currently published 
state-of-the-art. The theoretical interpretation of those measurements
have been widely discussed elsewhere \cite{RaWa,gaha}, however it is worthwhile
here to compare two approaches and to use this comparison to assess the 
control one has over the calculations. As a reminder, the rate of emission
of dileptons is related to the retarded in-medium photon
self-energy at finite temperature, $\Pi^{\rm R}_{\mu \nu} (E, q, T)$ \cite{ret}:
\begin{equation} 
E_+ E_- \frac{d^6 R_{\ell^+ \ell^-}}{d^3 p_+ d^3 p_-} = \frac{2 e^2}{(2 \pi)^6}\, 
\frac{n_{\rm B} (E,T)}{M^4}\, L^{\mu \nu} {\rm Im}
\Pi^{\rm R}_{\mu \nu} (E, p, T)
\label{eq1}
\end{equation}
where $n_{\rm B} (E, T)$ is a Bose-Einstein distribution function for energy
$E$ and temperature $T$, $M$ is the invariant mass of the lepton pair ($M^2 =
(p_+ + p_-)^2$), and
$L^{\mu \nu} = p^\mu_+ p^\nu_- + p^\mu_- p^\nu_+ - g^{\mu \nu}  p_+ \cdot
p_- $. 
A similar expression is derived for real photons. Owing to the phenomenological
success of vector meson dominance (VMD) \cite{sakurai}, the current-field identity
links the rate of electromagnetic emission directly to the in-medium vector
spectral density. It is therefore clear that measurements involving real or/and
virtual photons have the potential to reveal pristine features of the strongly
interacting many-body system. The electromagnetic emissivity can be calculated
in the hadronic sector by considering effective Lagrangians for the interacting
fields, and then by evaluating the vector spectral density \cite{RaWa,raga}. 
Another approach consists of using the relationship between the self-energy and
the forward scattering amplitude \cite{fscatt}, and by modeling the
latter by assuming that the dominant contributions are constituted of 
resonances coupled with a Pomeron background \cite{ebek}. There, the
forward scattering amplitude is then fitted directly to experimental data. Those
two approaches are of course related, but do constitute distinct avenues of
investigation of a common theme. The in-medium vector spectral densities
are computed, the rates evaluated with Eq. (\ref{eq1}), and the results are shown in 
Fig. \ref{rate_comp}. As a preamble, it is clear that the fitted scattering 
amplitudes can loose precision as the process moves further away from
on-shellness, but the two approaches clearly yield very similar dilepton 
production rates over the temperature and density range shown here. 
This speaks to the robustness of the theoretical results and both those
calculations contribute to the consensus of the need for modified in-medium
spectrum densities to explain low mass dilepton data at the SPS
\cite{cabrat,RaWa,brrho,gaha}. The specific nature of the
modification can't be singled out by the current experimental data, but the
importance of the enhancement at low energies is consistent with hadronic
many-body calculations. 

\section{Intermediate-mass lepton pairs}
\begin{figure}[!th]
\begin{center}
\resizebox{0.444\textwidth}{!}{%
  \includegraphics{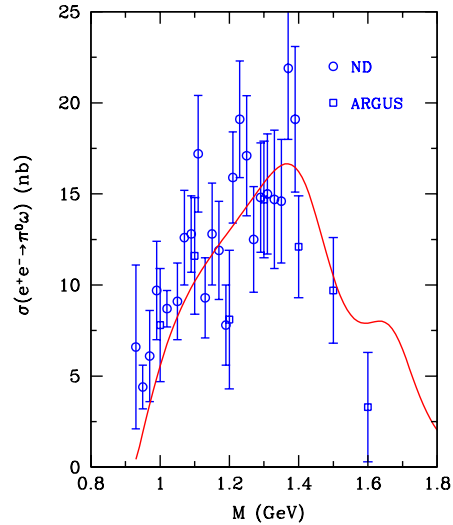}
}
\end{center}
\caption{The cross section for $e^+ e^- \to \omega \pi^0$. The data are from the
ND \cite{eedata} and ARGUS \cite{ARGUS} collaborations. The solid curve is generated
from a model described in \cite{eedata}.}
\label{piom}       % Give a unique label
\end{figure}

At intermediate mass ($m_\phi < M < m_{J/\psi}$), original estimates of the
dilepton production rate appeared especially promising, as kinematical
considerations combined with the original high
temperature of the QCD plasma would highlight the intermediate invariant mass
region as the window of opportunity for the observation of plasma radiation
\cite{shu78,k2m2}. 
Now, whether one uses effective
hadronic Lagrangian techniques or whether the self-energies are modeled directly
from the available empirical data, as discussed previously, the same problem
emerges. In both cases, the available parameters are fitted to measured 
physical properties which are softer than the scale defined by the intermediate 
lepton pair invariant mass. In this case, the appearance of off-shell effects 
is a genuine concern. Indeed, different approaches that agree in the soft
sector can yield widely different results in higher invariant mass extrapolations
\cite{gaogale}. Fortunately, constraints on the hadronic virtual photon-generating
processes can be obtained through the wealth of data of the type $e^+ e^- \to $
hadrons \cite{eedata}. Those measurements cover precisely the same invariant mass
range as the one that concerns us here. They have been used, together with
$\tau$-decay data, to construct the axial vector and vector spectral 
densities that are related to the
lepton-pair spectrum \cite{huang}. In addition, the intermediate-mass $e^+ e^-$
initial-state data have been analyzed specifically in a channel-by-channel
fashion. An example is shown in Fig.~\ref{piom}. 
This information can then be used to derive rates for hadrons $\to e^+ e^-$.
Following this procedure, the contributing channels for producing lepton pairs in the
appropriate invariant mass range are found to correspond to the initial states:
$\pi \pi$, $\pi \rho$, $\pi \omega$, $\eta \rho$, $\rho \rho$, $\pi a_1$, $K
\bar{K}$, $K \bar{K}^* + \rm{c.c.}$. A detailed discussion is too long to be had
here, but those channel are identified as the dominant ones, as their net lepton pair
contribution is found to saturate the spectral density analysis, at temperatures
relevant for the experimental measurements at hand \cite{kgs02}. With some
confidence in the microscopic rates, those can be integrated with an appropriate
modeling of the space-time evolution of the colliding system. 
%One approach that
%generates the appropriate hadronic phenomenology (hadronic spectra and their
%species-dependence, for example) is boost-invariant isentropic 
%hydrodynamics. There, matter
%undergoes a longitudinal expansion and an azimuthally-symmetric radial expansion,
%with a transition to a hot hadronic gas that consists of all hadrons having $M < $
%2.5 GeV, which makes for a rich equation of state. During the evolution, the speed
%of sound is calculated is consistently calculated at every temperature to be used
%in the equation of state needed for closing the hydrodynamic equations \cite{crs}. 

\begin{figure}[!th]
\begin{center}
\resizebox{0.444\textwidth}{!}{%
  \includegraphics{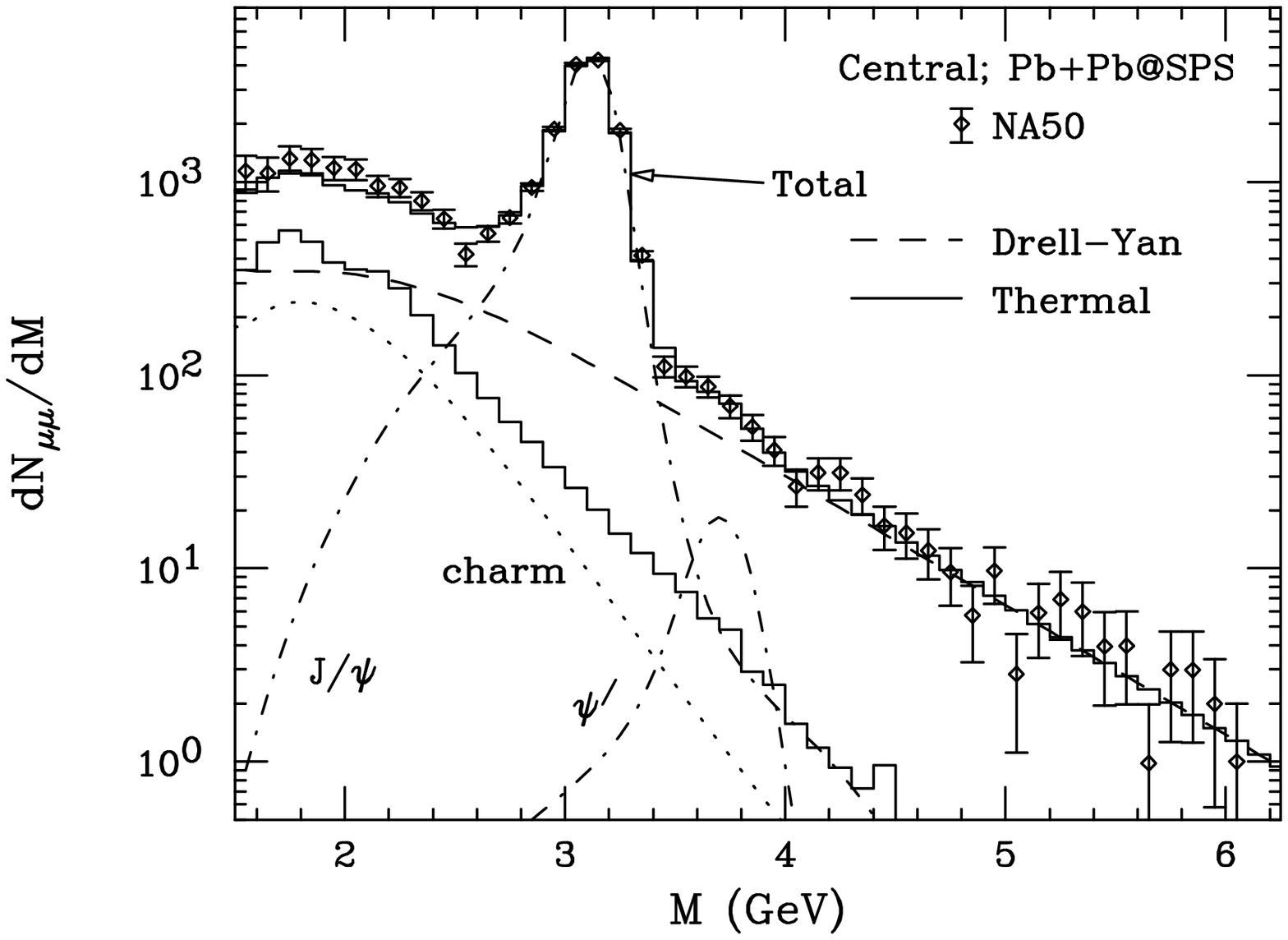}
}
\resizebox{0.444\textwidth}{!}{%
  \includegraphics{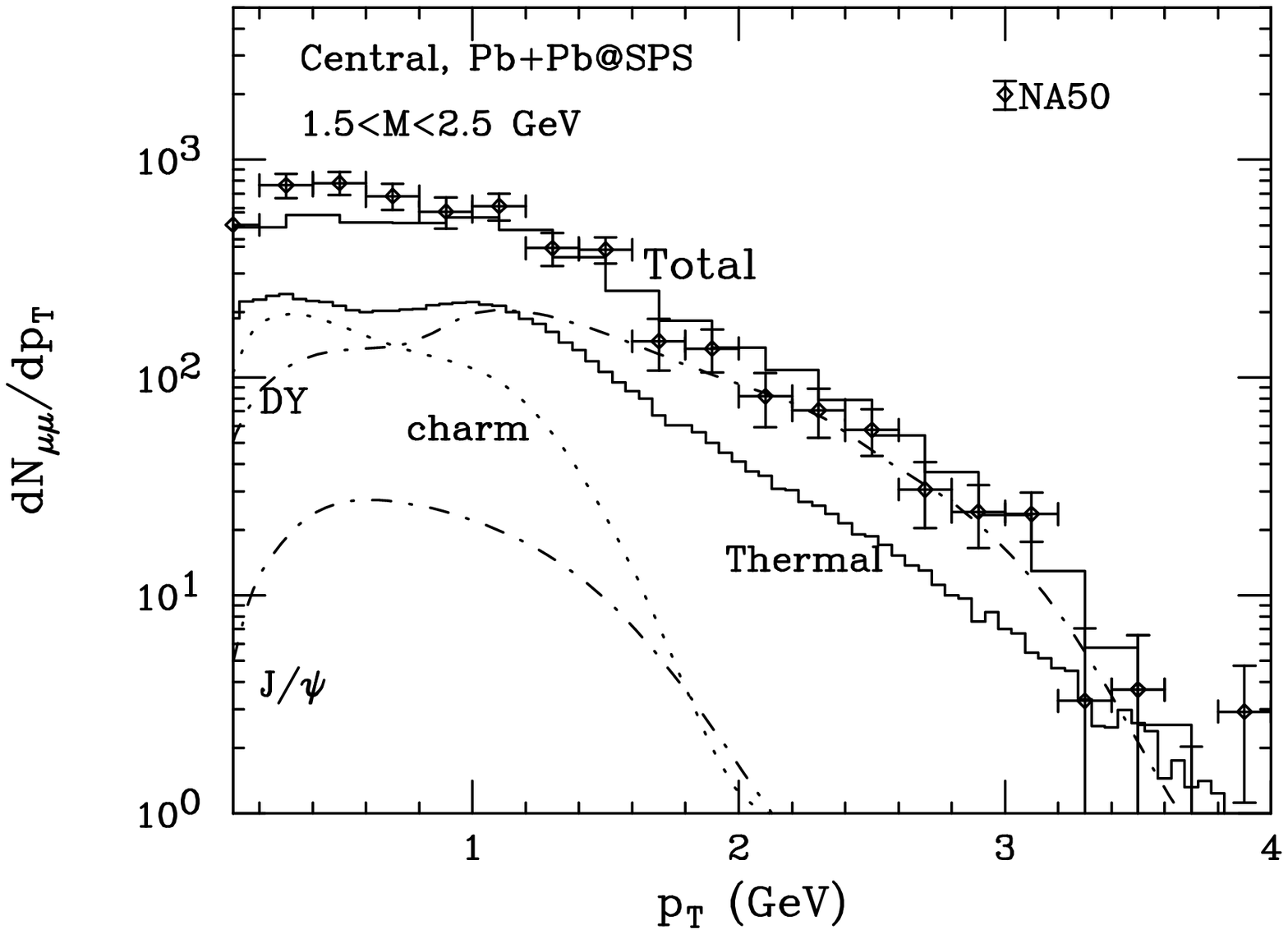}
}
\end{center}
\caption{The calculated dimuon invariant mass and transverse momentum spectrum. 
The sources are Drell-Yan, correlated charm decay, direct charmonium decay, and
thermal (quark-gluon plasma + hadron gas). The histogram represents the net
contribution, after correcting for detector acceptance, resolution, and efficiency.}
\label{imass}       % Give a unique label
\end{figure}
Experimentally, an excess of intermediate invariant-mass 
dimuons over those from sources expected from p-A
measurements has been confirmed by the Helios/3 \cite{he3} and NA50
collaborations \cite{na50}. We concentrate the latter. Note (as in most
experiments of this type) that it is important to
properly account for the detector's finite acceptance, as well as for its 
resolution and efficiency. A numerical filter has been developed specifically for this
purpose \cite{dra}. From the point of view of hard probes, this observed excess has
generated a fair amount of interest. Indeed, this invariant mass region is
sensitive to the irreducible background constituted by correlated open charm 
semileptonic decay \cite{shor}, and the excess can perhaps then be interpreted
either as an
increase in primordial $c \bar{c}$ abundances, or as a kinematical broadening of the
irreducible background generated by the rescattering of open charm mesons
\cite{linwang}. However, before more exotic explanations can be invoked, the
contribution of thermal meson sources needs to be assessed quantitatively. Similar
reasoning has been used in analyses of the Helios/3 \cite{ligale} and 
NA50 \cite{rashu} data.

Putting all of the elements described above together, we
arrive at the spectra shown in Fig.~\ref{imass}. The parameters that enter this
boost-invariant 
hydrodynamic calculation are the temperatures:  initial, critical and freezeout.
The set of those that is associated with Figure \ref{imass} is (330, 180, 120)
MeV. It is fair to say, however that the initial temperature determination is
somewhat dependent on the specific space-time modeling. 
However, a fairly robust conclusion still emerges: the intermediate
mass NA50 data does not demand a large radiation component from a plasma phase
(it is about 20\% here), nor does it require a large enhancement of the initial charm
content. Even though the dynamical models differ in detail, this bottom line is
shared by other studies of a similar nature \cite{ligale,rashu,gkp}. The new
high-precision data from NA60 \cite{NA60} is eagerly awaited.

\section{Low $p_T$ photons}

At SPS energies, real photon spectra  have been measured by the WA98 collaboration
\cite{WA98}. Those data have been interpreted within several different approaches,
such as hydrodynamic simulations \cite{hydro},  transport/cascade simulations  
\cite{casc}, as well as using simple fireball models \cite{fire}. We describe here
a recent calculation where the microscopic rates have been revisited, with an
emphasis put on basic hadronic phenomenology.  We have described already the
connection between the photon production rate and the in-medium vector spectral
density. For self-energy topologies up to two loops, the imaginary
part is readily shown to reduce to tree-level diagrams, in which case a 
kinetic theory approach proves to be convenient. In such a framework
\begin{eqnarray}
E \frac{d^3 R}{d^3 q} = \int \frac{d^3 p_1}{(2 \pi)^2 2 E_1} 
\frac{d^3 p_2}{(2 \pi)^2 2 E_2}\frac{d^3 p_3}{(2 \pi)^2 2 E_3} 
(2 \pi)^4 \mbox{\ \ \ \ \ } \nonumber \\
\times \, \delta^{(4)} ( p_1 + p_2 - p_3 - q) | {\cal M} |^2 \frac{f(E_1) f(E_2) [1 \pm
f(E_3)]}{2 (2 \pi)^3}
\end{eqnarray}

\begin{figure}[!th]
%\begin{center}
\resizebox{0.544\textwidth}{!}{%
  \includegraphics[angle=-90]{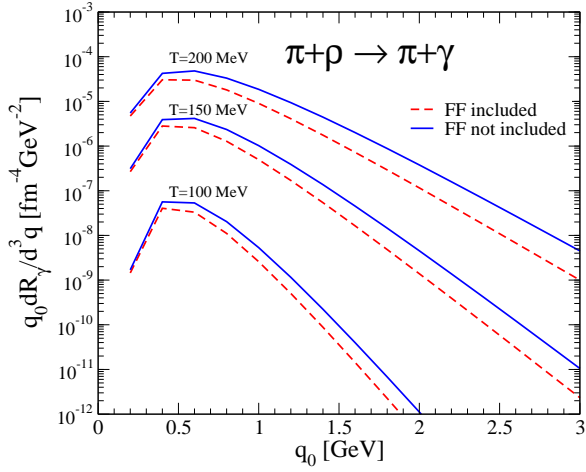}
}
%\end{center}
\caption{Photon production rates at three different temperatures, showing the
effect of form factors at hadronic vertices.}
\label{formf}       % Give a unique label
\end{figure}

\begin{figure}[!th]
%\begin{center}
\resizebox{0.444\textwidth}{!}{%
  \includegraphics[angle=-90]{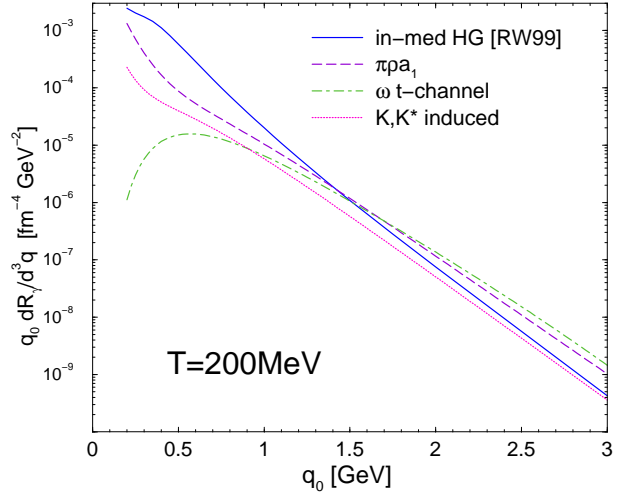}
}
%\end{center}
\caption{Comparison of different sources for photon production in a hot and dense
hadronic gas with $T$ = 200 MeV, and baryon chemical potential $\mu_{\rm B}$ = 220
MeV. The dashed and dotted curves represent the photon rates calculated in the MYM
approach without the $t$-channel $\omega$ exchange. This latter contribution is
shown by the dashed-dotted line. The full curve is the photon emissivity obtained
with the vector spectral function approach including baryons.}
\label{mesrates}       % Give a unique label
\end{figure}

Considering first the baryon-free sector, the elementary photon-producing
reactions that involve light pseudoscalars, vectors, and axial vector mesons are
evaluated in the massive Yang-Mills (MYM) formalism. This framework is capable of
describing adequate hadronic phenomenology with a limited set of adjustable
parameters \cite{srg}. The vector and axial vector fields are treated as massive
gauge fields of the chiral U(3)$_L$ $\times$ U(3)$_R$ symmetry, and added to the
nonlinear $\sigma$ model formulated in the exponential representation \cite{gks}. 
Note that this form of the interaction permits a coherent treatment of the strange
and nonstrange sectors of the theory, and thus does not suffer from phase
ambiguities. Proceeding further, an expansion of the
Lagrangian enables a systematic evaluation of all relevant processes. More
specifically, all Born-level graphs with the appropriate crossing-symmetry
partners were considered for reaction- and decay-type processes. An important
consideration in applying effective hadronic models at moderate and high 
momentum transfers is the use of hadronic form factors. Those arise generally 
in effective
models and are ubiquitous in hadronic physics. They are incorporated in the fits
to hadronic properties, consistently with electromagnetic current conservation 
requirements \cite{srg}. Their effect at different temperatures may be judged from
Fig. \ref{formf}. 
An additional point worth mentioning in this context is that the $\omega$ vector
meson is known to exhibit a large coupling to $\pi \rho$, and thus to $\pi
\gamma$, owing  to VMD. The on-shell radiative decay contribution is included
in the early estimates of photon production \cite{kls}, but its $t$-channel
exchange in the reaction $\pi \rho \to \pi \gamma$ has not received much attention
up to now. However, the usage of the hadronic form factors forces a re-calibration
of the coupling constants. Because the $\omega \rho \pi$ vertex is constrained by
the radiative decay width of the $\omega$, and because this decay process involves
an off-shell hadronic vertex (owing to VMD), the coupling is modified by the
presence of the form factor. The size of this specific contribution can be
assessed by considering the information in Fig. \ref{mesrates}. 

\begin{figure}[!th]
\resizebox{0.444\textwidth}{!}{%
  \includegraphics[angle=-90]{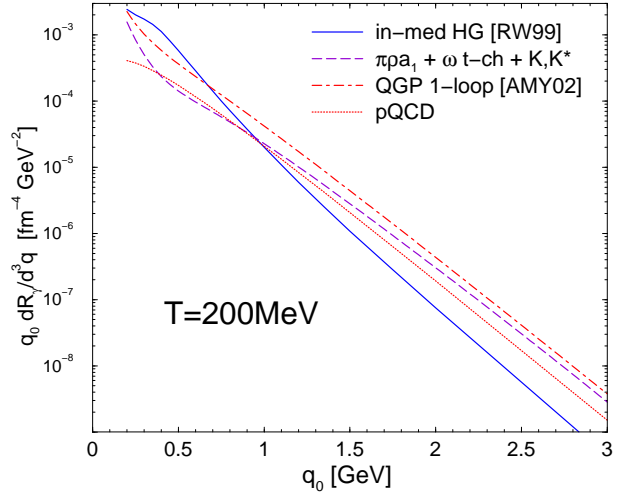}
}
\caption{Comparison of rates for photon production from a hot gas of partons. The
solid line is from the many-body approach of Ref. \cite{RaWa}, the dashed line
represent the mesonic contributions, the dotted line is the HTL-corrected pQCD
result, and the dashed-dotted line is the result that is complete at leading order
in $g_s$.} 
\label{QCDratecomp}       % Give a unique label
\end{figure}

Since the emission of lepton pairs and that of real photons are linked to the
same object, the in-medium photon self-energy, both should be calculable within
the same formalism. This is what is done in the work we describe. Care has to be
taken, as the leading order contribution in both cases belong to different
self-energy topologies. Moreover, the issues of double counting and coherence have
to be considered. The $a_1$ $s$-channel graph is present in both the $\rho$
spectral density and in the MYM framework. We remove it from the former, where it
plays a minor role, whereas it induces non-negligible interference effects in the
$\pi \rho a_1$ complex. If coherence is not important, the $t$-channel
contributions may be evaluated separately. It was explicitly verified that this
was the case for the $\omega$ exchange. The photon rate induced by bringing the
vector spectral density to the photon point is shown in Fig. \ref{mesrates} by the
solid line. It is instructive to compare the hadronic photon emission rates with
those from a hot gas of partons at a similar temperature. This is done in Fig.
\ref{QCDratecomp}. 
There, the spectral strength of the meson sources is compared with that in the
baryon-rich sector, at a temperature of 200 MeV. Also shown: the 
hard-thermal-loop-corrected (HTL) result \cite{kls,phot} (labeled pQCD), and the
complete leading-order in $g_s$ result for the photon emissivity of the
quark-gluon plasma \cite{AMY}. It would be useful to extract the required
spectral density from lattice QCD calculations, but efforts there are 
beginning \cite{latt}. 

\begin{figure}[!th]
%\begin{center}
\resizebox{0.444\textwidth}{!}{%
  \includegraphics[angle=-90]{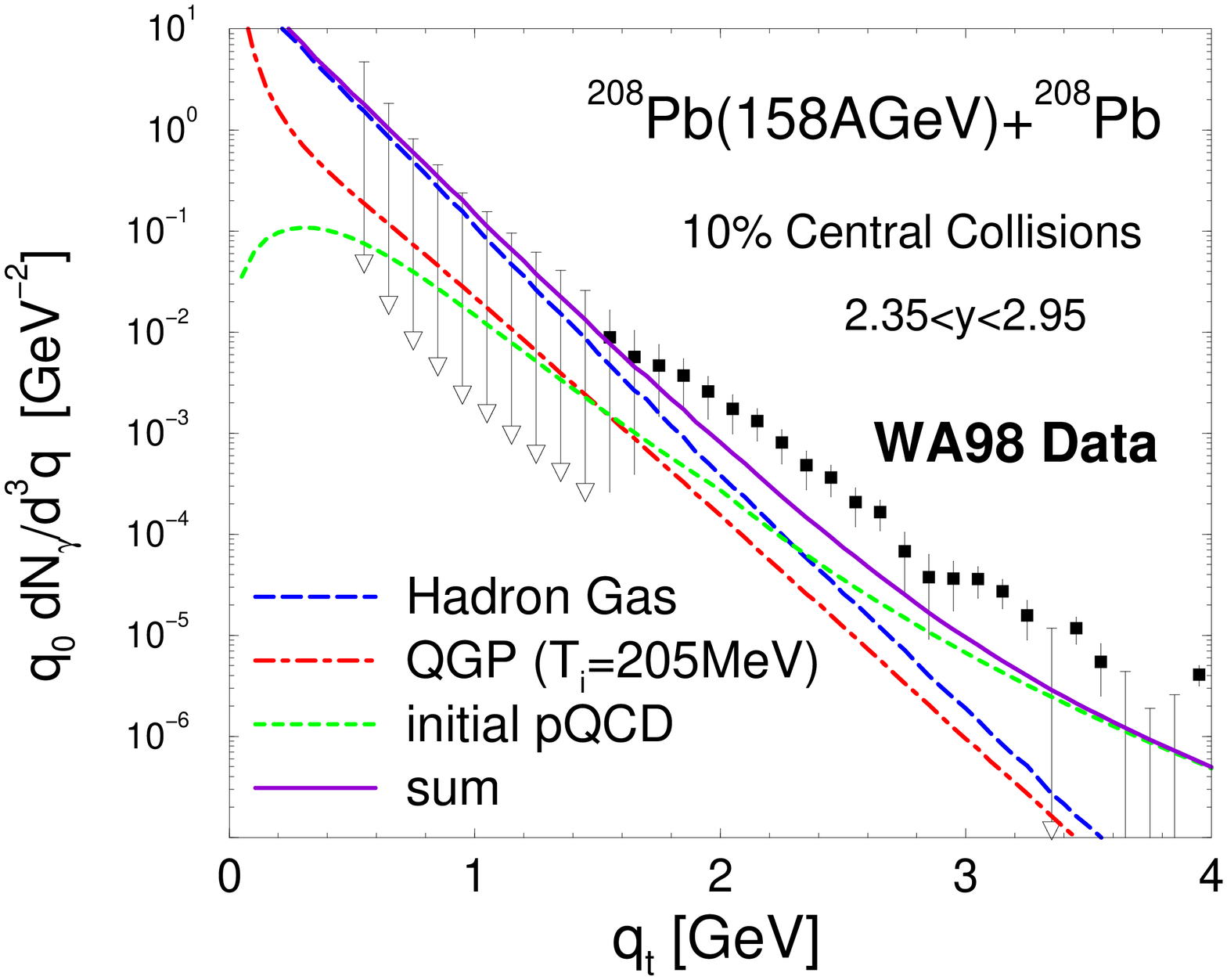}
}
\resizebox{0.444\textwidth}{!}{%
  \includegraphics[angle=-90]{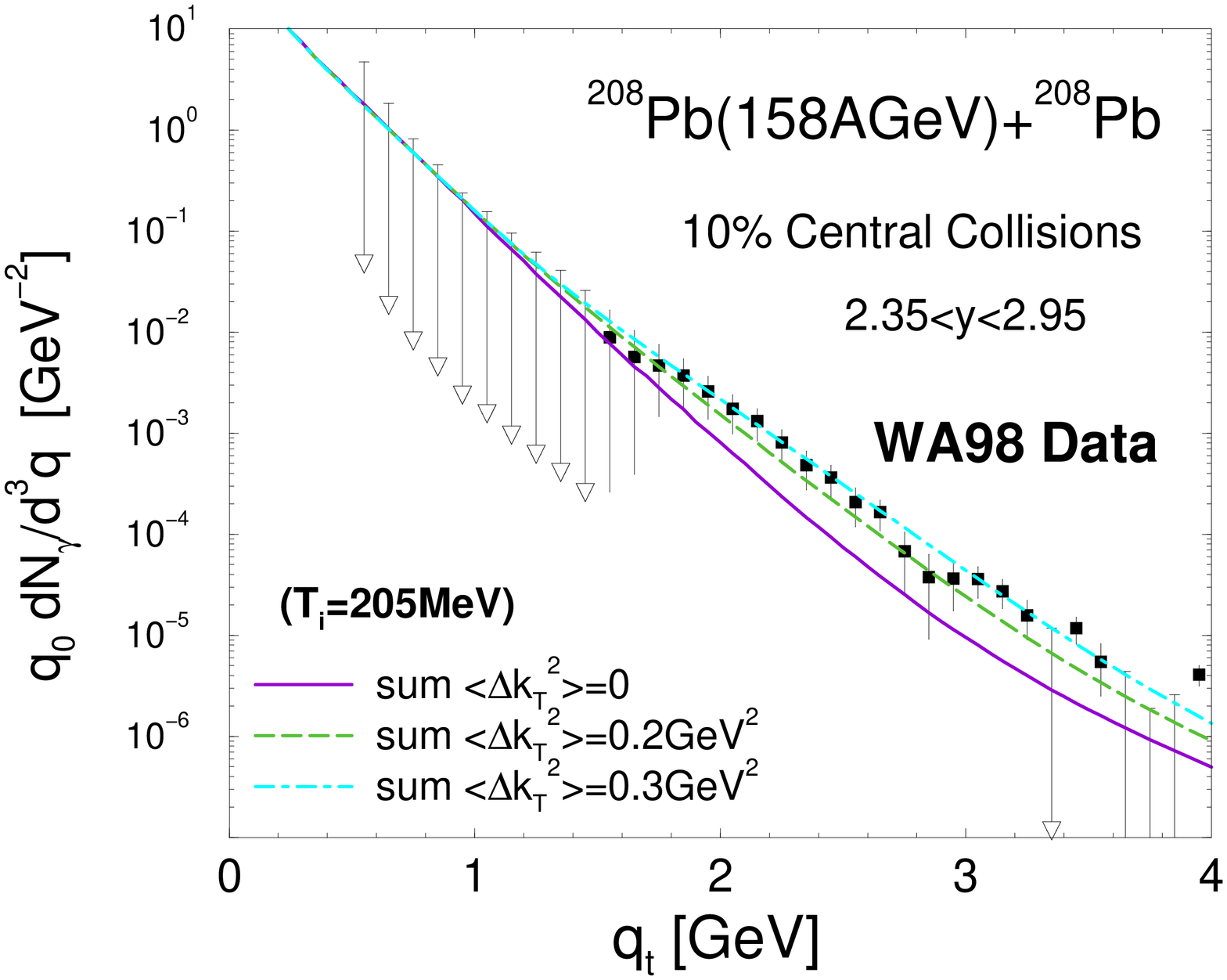}
}
%\end{center}
\caption{Top panel: Thermal plus prompt photon spectra compared to data from
WA98, for central Pb + Pb collisions at the SPS. Lower panel: the effect of the
nuclear transverse momentum broadening on the measured photon spectrum. In
analyses of p-A photon data, an adequate reproduction of the appropriate
measurements emerges with a value of the broadening parameter $\langle \Delta
k_T^2 \rangle \simeq 0.1 - 0.2$ GeV$^2$. }
\label{wa98}       % Give a unique label
\end{figure}

Additional aspects need to be discussed before final yields can be derived.
The emission rates again need to be integrated over the space-time history of 
the collision event. This is done here with a fireball model, which incorporates the
main elements of hydrodynamic calculations. Soft photons are associated with
sources which emit late in the space-time history of the reactions, and are thus
sensitive to details of the flow profile. The details appear elsewhere
\cite{RaWa2,srg}, but using conservation laws one is able to extract the temperature and
baryon chemical potential at any proper time, and to define a trajectory in the $\mu_{\rm
B} - T$ plane. The transition from the plasma to the hadronic gas phase is set at
the chemical freeze-out locus experimentally extracted from hadron species ratios
\cite{pbm}. The hadronic gas is then evolved from chemical to thermal freeze-out by
introducing appropriate chemical potentials. The local photon momentum
distributions are finally boosted to the lab frame, according to the
time-dependent transverse expansion velocity that is eventually also found in the
measured hadron transverse spectra. In addition, contributions to the direct
photon spectra come from prompt photons emitted in primordial nucleon-nucleon
collisions. An accurate theoretical description thereof at SPS energies 
is still a matter of
debate \cite{aur}, therefore an empirical scaling relationship extracted 
from fits to data \cite{dks} is used. Finally, the transverse momentum
broadening generated by the nuclear medium (Cronin effect) is estimated from
analyses of p-A data \cite{srg}. The result of considering all of the
aspects discussed up to now, together with an adequate dynamical 
modeling of the nuclear collisions for the WA98 experiment appears on Fig.
\ref{wa98}.
The initial temperature used here is part of a global analysis of low and
intermediate mass lepton pair spectra \cite{RaWa2,rashu}. 

A partial summary is
possible and appropriate here. As the primordial
microscopic rates are very similar, it is increasingly clear that the
differences in some of the intrinsic parameters of the various theoretical
analyses, such as temperature, are related to differences in the space-time
evolution. Nevertheless, the robust features here are that intermediate mass
lepton pair spectra, as well as low mass dilepton and 
real photon spectra can be understood
in terms of hadronic degrees of freedom. Furthermore, 
low mass dileptons and low $p_T$ 
real photons
are consistently calculated with the same spectral densities. 
The quark-gluon plasma component in all
cases is not considerable enough to permit an unambiguous identification.
For RHIC and the LHC, however, the situation is more promising
\cite{srg}, as we now discuss.

\section{High $p_T$ photons and jets}

\begin{figure}[!th]
\resizebox{0.444\textwidth}{!}{%
  \includegraphics[angle=-90]{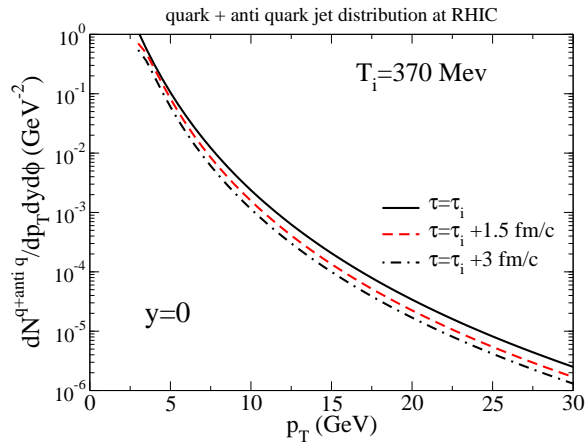}
}
\caption{The time-development of a quark (or antiquark) jet 
momentum-profile, in an evolving parton medium. 
The initial
distribution is obtained from a NLO calculation \cite{aurenche} that includes
initial-state shadowing \cite{esko}, for RHIC conditions. }
\label{jet-evol}       % Give a unique label
\end{figure}

\begin{figure}[!th]
\resizebox{0.444\textwidth}{!}{%
  \includegraphics[angle=-90]{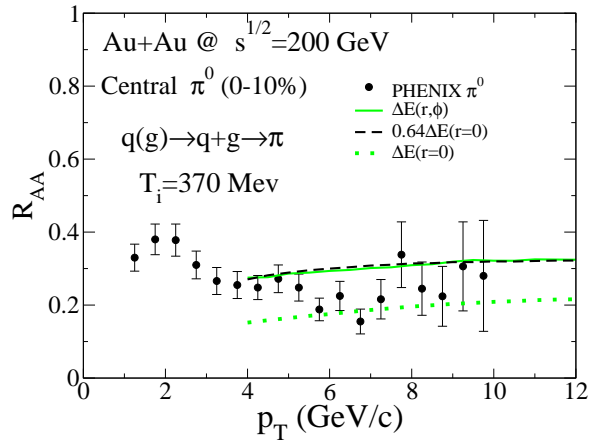}
}
\caption{$R_{AA}$ for neutral pions measured by the PHENIX experimental
collaboration. See text for details.} 
\label{raa}       % Give a unique label
\end{figure}

One of the most striking discoveries at RHIC has been that of the disappearance
of hadron-hadron correlation \cite{corr} and of the suppression of 
single-particle spectra in
central nuclear collisions \cite{jets}. A compelling theoretical interpretation of those
results is that of jet absorption in hot and dense partonic matter, signaling in
effect the existence of a quark-gluon plasma. Several models of jet-quenching 
through gluon bremsstrahlung have been elaborated \cite{bdmps,ww,glv,kw,z}. 
Here, we shall report on results obtained using the approach developed by
Arnold, Moore, and Yaffe (AMY) \cite{AMY1}. There, the  initial hard gluon
($P_g (p, t=0)$) and
hard quark plus antiquark ($P_{q \bar{q}} (p, t=0)$) probability 
distributions are evolved with time, as
they traverse the medium. The joint equations for those quantities can be
visualized as Fokker-Planck equations \cite{tgjm}. This technology permits,
given an initial jet profile calculated from zero-temperature QCD, 
to visualize its time-evolution, as shown in Fig. \ref{jet-evol}. 
One may then investigate the effect of energy loss  (and gain)  on hadronic 
and electromagnetic observables.  As mentioned previously, one 
variable that is often
invoked in the context of jet quenching discussions, is that associated with the
suppression of single-particle momentum distributions. A quantitative measure
of this suppression is shown in the so-called $R_{A A}$ profile, where 
\begin{eqnarray}
R_{AA} = \frac{dN_{AA}/dy d^2 p_T}{\langle N_{\rm coll}\rangle dN_{pp}/dy 
d^2 p_T} 
\end{eqnarray}
is plotted as a function of transverse momentum. 
Clearly, if a nucleus-nucleus collisions is nothing  more than a superposition
of nucleon-nucleon collisions, then $R_{AA}$ should be unity. The main points in 
the calculation of this quantity may be summarized here. First, in dense matter 
the parton distribution functions
are different than what they are in proton-proton collisions \cite{esko}. Also,
it is assumed that  a jet fragments outside the strongly interacting medium,
as suggested by formation time arguments, and that the fragmentation involves
vacuum fragmentation functions. The effect of the medium is then to
reduce the parton energy by an amount determined by the time evolution of the
energy profile shown, for example, in Fig. \ref{jet-evol}. The jet
starts in the QGP medium and evolves until it reaches the surface, or until
the medium reaches the transition temperature, $T_c$. Note that we assumed
that, at early times, the plasma could be modeled as following an 
isentropic 1-D evolution, and that a first-order phase transition exists with
a critical temperature of 160 MeV. Finally, it is important to point out that
the spectrum calculated without energy loss is completely in agreement with
measurements done in proton-proton collisions \cite{tgjm}. 
Note that since jets are emitted early in the collisions, the final profile
shows only modest sensitivity to details of the time-evolution \cite{tgjm}. 
For Au-Au central
collisions at RHIC energy, we obtain the $\pi^0$ results shown in Fig.
\ref{raa}. 
We also assume a realistic spatial distribution for the jet initial location.
Then, the neutral pion spectrum is obtained and shown by the full curve. If
one makes the simplifying assumption that all jets originate from the centre
of the nucleus (what is not done in the rest of this work), one obtains the 
lower dashed curve. The third line shows that
one is related to the other by a constant, up to a very good
approximation. Within the formalism of AMY, the only explicit parameter in this
calculation that is not common to other phenomenological studies of 
RHIC results (both hadronic and electromagnetic, see for example Ref.
\cite{srg}) is the strong coupling constant, $\alpha_s$. We use $\alpha_s$ =
0.3. 

\begin{figure}[!th]
\resizebox{0.444\textwidth}{!}{%
  \includegraphics[angle=-90]{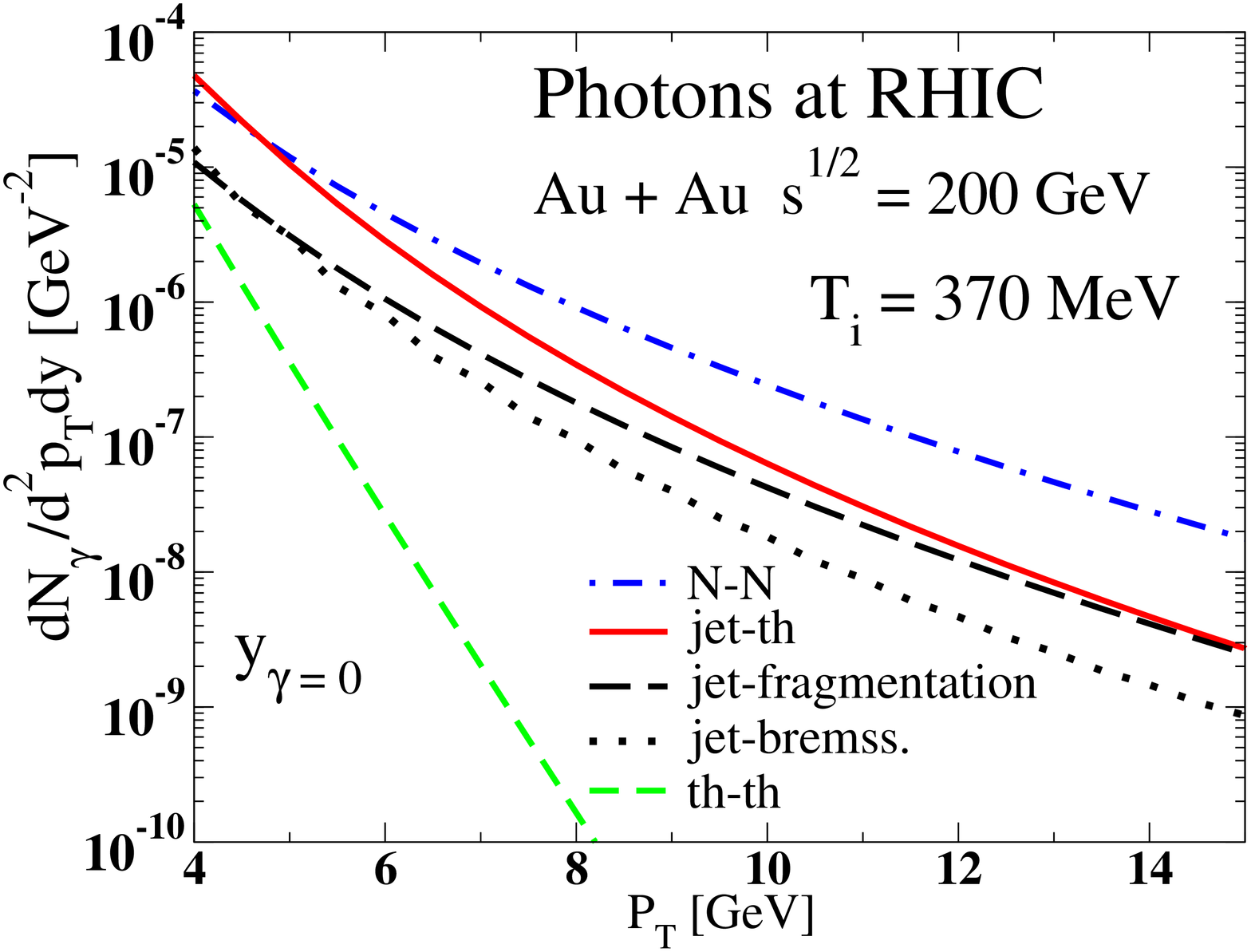}
}
\resizebox{0.444\textwidth}{!}{%
  \includegraphics[angle=-90]{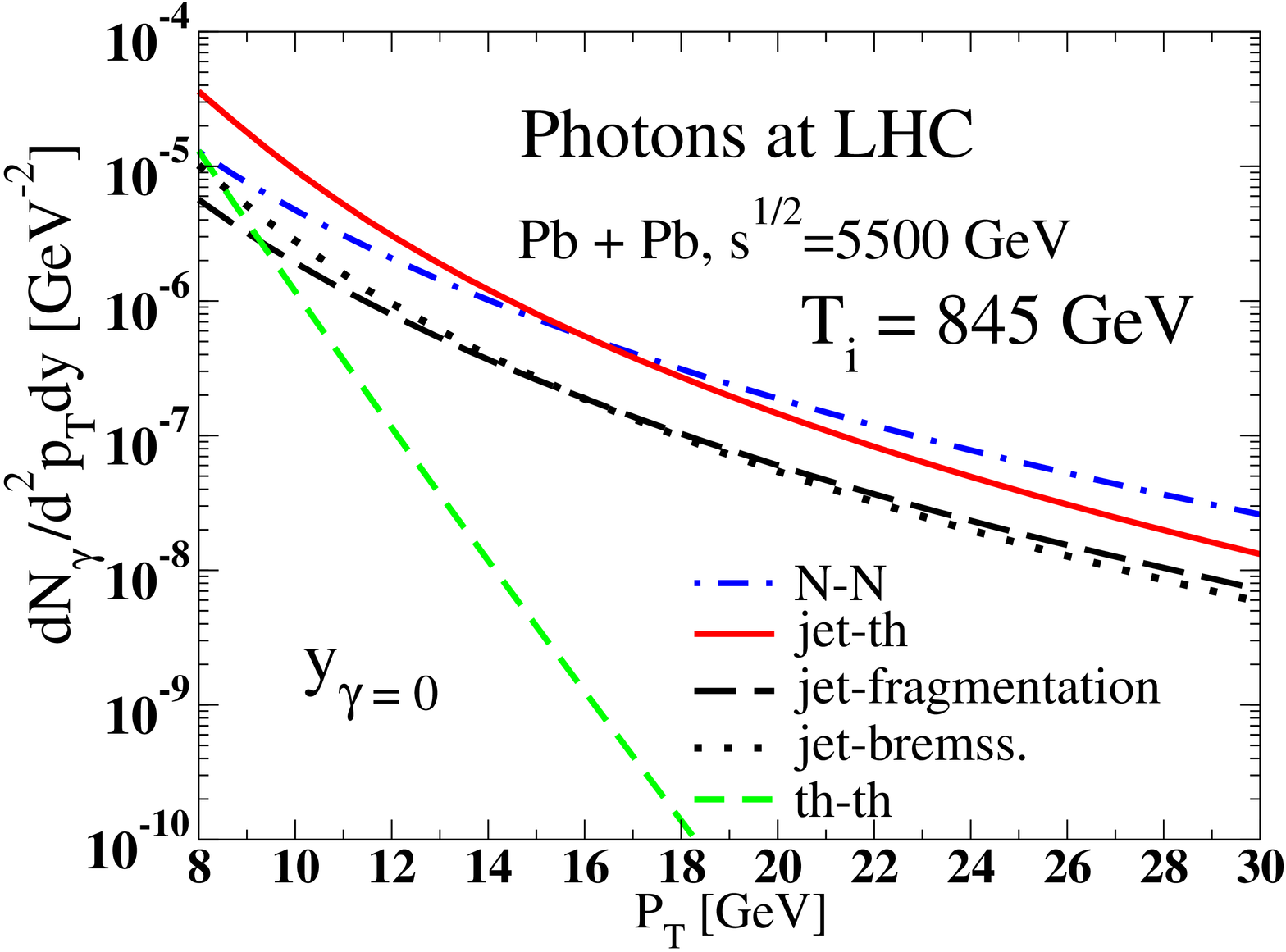}
}
\caption{Sources of high $p_T$, mid-rapidity photons in central Au-Au
collisions at RHIC (top panel), and at the LHC (lower panel). The different
sources are discussed in the text.}
\label{phot_sources}       % Give a unique label
\end{figure}

\begin{figure}[!th]
\resizebox{0.444\textwidth}{!}{%
  \includegraphics[angle=-90]{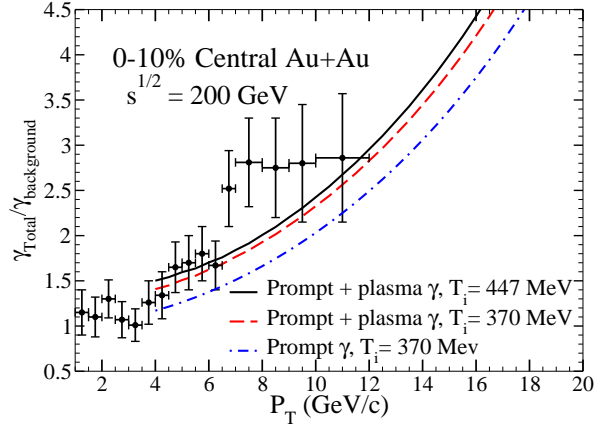}
}
\caption{The ration of all photons over the decay photons is shown, for Au-Au
collisions at RHIC energies, with and without the thermal contribution. The
effect of varying the initial temperature is also shown. The data are from the
PHENIX collaboration \cite{phenix}.}
\label{gambck}       % Give a unique label
\end{figure}

If the physical conditions for jet quenching are realized, they do signal 
a jet-plasma interaction. By the same argument, this interaction  can 
manifest itself through other probes, some of which may be 
electromagnetic. Previous estimates have shown that the conversion of 
a leading parton to a photon in the plasma was found to be an important 
source of real photons \cite{fms}. This means that a jet crossing the hot
medium undergoes an annihilation ($ q + \bar{q} \to g + \gamma$) or a Compton
process ($g + q \to q + \gamma$) with a thermal parton. The contribution to
the photon production rate in a finite-temperature parton medium for the
leading topology is known to be \cite{phot,fms}
\begin{eqnarray}
\frac{d R}{d y d^2 p_T} = \sum_f \left(\frac{e_f}{e}\right) \frac{T^2 \alpha
\alpha_s}{8 \pi^2} \left[ f_q (\vec{p}_\gamma) + f_{\bar{q}}
(\vec{p}_\gamma)\right]\nonumber \\
\times \left[ 2 \ln \left(\frac{4 E_\gamma T}{m^2}\right) -
C_{\rm ann} - C_{\rm Comp}\right]
\end{eqnarray}
where $T$ is the temperature, $C_{\rm ann}$ = 1.916, and $C_{\rm Comp}$ =
0.416. In a hot QCD medium, the infrared singularity that appears in the limit
of vanishing quark mass, $m \to 0$, gets screened by hard thermal loops: $m^2 = 4
\pi \alpha_s T^2/3$ \cite{kls,phot}. The incoming parton may now be the leading
parton in a jet, and then strikes a thermal parton. However, the jet evolving
in the QCD medium has lost energy and this is accounted for with the
technology described earlier. We term this photon source ``jet-thermal'', in an
obvious nomenclature. The net photon spectrum will also receive contributions
from sources identified with primordial hard nucleon-nucleon collisions, with
the jet fragmenting into a photon (and hadrons, after loosing energy), with
the jet producing photons via bremsstrahlung interactions as it traverses the
medium (and looses energy), and with photons produced through 
interactions of thermal plasma constituents \cite{tgjm}. The emissivity in 
those different channels is integrated over the space-time history, with initial conditions appropriate for
RHIC and the LHC, and the result is shown in Fig. \ref{phot_sources}. 
Both at RHIC and LHC energies, it is satisfying to note that the original
premise of this exercise still holds true: the jet-plasma photons are an
important source, which in fact outshines others 
at $p_T \sim$ 4 GeV/c
for RHIC, and at $p_T \sim$ 8 GeV/c for the LHC. At RHIC, real photon data
already 
exists and there is much more to come. An early analysis concentrated on the
ratio of the total number of photons to background photons:
\begin{eqnarray}
\frac{\gamma_{\rm total}}{\gamma_{\rm background}} = 
\frac{d^3 N_{\gamma}^{\rm bck}/d^2 p_T d y + \sum {\rm all\ other\ sources}}{d^3
N_{\gamma}^{\rm bck}/d^2 p_T dy}
\end{eqnarray}
This quantity is plotted in Fig. \ref{gambck}, together with data from PHENIX
\cite{phenix}, with and without the thermal contribution. The calculation
including the thermal component is in good agreement with the data, except for
a few points in the range $7 < p_T < 9$ GeV/c. Without the thermal components
the agreement worsens considerably. The small effect of varying the initial
temperature is also seen on the same figure. 

\section{Summary}
Soft electromagnetic spectra (low mass dileptons and low $p_T$ photons) receive
an important contribution from hadronic sources. We have shown results where
the emissivity has been derived from the same in-medium spectral density and
continued to the time-like sector and to the light cone, respectively, and
convolved with the same dynamical model. It is fair to say that the SPS data does not demand 
a quark gluon plasma contribution, in a direct manner. This statement also holds
true for the intermediate mass dileptons, where the radiation from thermal meson
channels was also found to be quantitatively important. At RHIC and LHC energies,  
a complete leading-order treatment of jet energy loss in the QCD plasma has been
used to calculate both pion and photon spectra. The
results have been confronted with RHIC data and turn out to be in good
agreement. This lends further support to the idea that high $p_T$ 
suppression, for the set of 
kinematical conditions considered here,  is a final-state effect mostly driven
by gluon bremsstrahlung in the hot medium.

\section*{Acknowledgments} 
It is a pleasure to thank my collaborators on much of the work presented
here: Sangyong Jeon, Ioulia Kvasnikova, Guy D. Moore, Ralf Rapp, Dinesh K. Srivastava, 
and Simon Turbide. I also 
thank Simon Turbide for a critical reading of this paper.  This work was
supported in part by the Natural Sciences and Engineering Research Council of
Canada, and in part by the Fonds Nature et Technologies of Quebec. I am grateful
to the Service de Physique Th\'eorique, CEA/Saclay, for providing facilities and
support used in the completion of this work.

%\begin{figure*}
%\vspace*{5cm}       % Give the correct figure height in cm
%\end{figure*}
%\begin{table}
%\begin{tabular}{lll}
%\hline\noalign{\smallskip}
%\end{tabular}
% Or use
%\vspace*{5cm}  % with the correct table height
%\end{table}
%
% BibTeX users please use
% \bibliographystyle{}
% \bibliography{}

\end{document}